\documentclass[11pt]{article}
\pdfoutput=1

\usepackage{euscript}
\usepackage{amssymb}
\usepackage{amsfonts}
\usepackage{amsbsy}
\usepackage{amsmath}
\usepackage{epsfig}
\usepackage{amsthm}
\usepackage{amscd}
\usepackage{amstext}
\usepackage{verbatim}
\usepackage[T1]{fontenc}

\def\ben{\begin{equation}}
\def\een{\end{equation}}

\let\a=\alpha  \let\g=\gamma  \let\e=\varepsilon

\let\s=\sigma

 \let\G=\Gamma

\let\pa=\partial
\def\be{\begin{equation}}
\def\ee{\end{equation}}
\def\beq{\begin{equation}}
\def\eeq{\end{equation}}
\def\ba{\begin{array}}
\def\ea{\end{array}}
\def\del{\partial}

\def\dalemb#1#2{{\vbox{\hrule height .#2pt
       \hbox{\vrule width.#2pt height#1pt \kern#1pt
               \vrule width.#2pt}
       \hrule height.#2pt}}}

\newcommand{\bea}{\begin{eqnarray}}
\newcommand{\eea}{\end{eqnarray}}

\def\Lag{{\mathcal{L}}}

\begin{document}

\begin{center}

{ \LARGE {\bf Holographic order parameter for \\
charge fractionalization}}

\vspace{1cm}

{\large Sean A. Hartnoll and \DJ or\dj e Radi\v cevi\' c
\vspace{0.7cm}

{\it Department of Physics, Stanford University, \\
Stanford, CA 94305-4060, USA \\}}

\vspace{1.6cm}

\end{center}

\begin{abstract}

Nonlocal order parameters for deconfinement, such as the entanglement entropy and Wilson loops, depend on spatial surfaces $\Sigma$. These observables are given holographically by the area of a certain bulk spatial surface $\Gamma$ ending on $\Sigma$. At finite charge density it is natural to consider the electric flux through the bulk surface $\Gamma$ in addition to its area. We show that this flux provides a refined order parameter that can distinguish `fractionalized' phases, with charged horizons, from what we term `cohesive' phases, with charged matter in the bulk. Fractionalization leads to a volume law for the flux through the surface, the flux for deconfined but cohesive phases is between a boundary and a volume law, while finite density confined phases have vanishing flux through the surface. We suggest two possible field theoretical interpretations for this order parameter. The first is as information extracted from the large $N$ reduced density matrix associated to $\Sigma$. The second is as surface operators dual to polarized bulk `D-branes', carrying an electric dipole moment.

\end{abstract}

\pagebreak
\setcounter{page}{1}

\section{Holographic phases of matter}

Certain large $N$ gauge theories admit a holographically dual description in terms of a theory of classical gravity in one higher spatial dimension \cite{Maldacena:1997re}. Many of the most interesting questions one would like to ask of such theories concern the gauge field dynamics. For instance, one would like to gain a dual geometrical understanding of confinement. This was largely achieved shortly after the discovery of holography \cite{Witten:1998zw}, as we will briefly review.

If the large $N$ gauge theory has more structure, one can ask more refined questions. In particular, if the gauge theory has a global $U(1)$ symmetry, then one can place the theory at a nonzero chemical potential $\mu$, which will typically induce a charge density $\langle J^t \rangle$. In this finite density state, one can ask whether the charge carriers are ``confined'' or ``deconfined''. As we shall elaborate throughout this paper, this is not the same notion of confinement and deconfinement as we have used in the previous paragraph, which referred to the charge-neutral glue sector. New concepts are required. Phases in which the charge carriers are deconfined are known as fractionalized phases of matter.\footnote{``Fractionalized'' is often used synonymously to ``deconfined'' in a condensed matter context, and also used to describe a class of gapped phases, but we will use the term more restrictively.}
We will introduce the notion of a ``cohesive'' phase to describe the opposite of a fractionalized phase, i.e.~when the charge carriers are confined.\footnote{The term ``mesonic'' has previously been used by one of the authors to describe the opposite of fractioanlization, but ultimately this may have misleading associations.}

Various (nonlocal) field theoretic order parameters exist for deconfinement. Confining and deconfined phases can be distinguished using Polyakov loops, Wilson loops, and the entanglement entropy. All of these have beautifully simple realizations in holography as certain surfaces in the bulk geometry \cite{Witten:1998zw, Maldacena:1998im, Rey:1998ik, Ryu:2006bv}. Deconfined phases are characterized by an event horizon in the bulk, while confined phases are dual to geometries that are effectively cut off at some IR scale. Similarly sharp order parameters do not yet exist for fractionalization, as distinct from deconfinement. The objective of this paper is to start to remedy this situation from a holographic perspective.

It has been proposed that fractionalization can be diagnosed by a mismatch in the Luttinger count in phases that do not break the $U(1)$ symmetry spontaneously \cite{fract,Sachdev:2010um, Huijse:2011hp}. This suggestion led directly to the observation that finite density holographic duals are naturally classified by whether the asymptotic electric flux emanates from behind an event horizon or is carried by charged matter in the bulk, e.g. \cite{Hartnoll:2011fn}. For fermionic bulk matter, the flux from behind the horizon indeed appears as a deficit in the Luttinger count. An elegant general proof of this statement was obtained in \cite{Iqbal:2011bf}, building on previous partial results \cite{Hartnoll:2011fn, Hartnoll:2011dm, Iqbal:2011in, Sachdev:2011ze}. Furthermore, this picture has lead to the interesting suggestion that in the superfluid case, where the $U(1)$ is spontaneously broken, fractionalization could be detected as a deficit in the transverse Magnus force acting on a moving superfluid vortex \cite{Iqbal:2011bf}.

It is desirable, however, to have a clean field-theoretic order parameter characterizing fractionalization, in the spirit of the various nonlocal order parameters defining deconfinement. In a holographic context, following the logic of the previous paragraph, this translates into finding a quantity that can be measured at the conformal boundary of the dual spacetime, and that detects whether the asymptotic electric flux originates from behind a horizon or not. In this paper we will construct two, closely related, such quantities. We do not know at this point precisely what the objects we consider correspond to in the boundary quantum field theory. This is clearly an important question. We will argue that they correspond in the first case to a refinement of the entanglement entropy that can be defined in systems at a finite charge density, and in the second case to operators describing branes in the bulk carrying a dipole moment. Let us first recall relevant established results concerning confinement in the neutral `glue' sector.

\subsection{Neutral sector: Confinement and deconfinement}

We will briefly review how confinement and deconfinement appear holographically. We will focus on the entanglement entropy. The entanglement entropy is associated with a spatial hypersurface $\Sigma_{d-1}$ in the $d+1$ dimensional field theory. 

The holographic prescription for calculating entanglement entropy has been proposed by Ryu and Takayanagi  \cite{Ryu:2006bv}: given a boundary hypersurface $\Sigma$, find the minimal surface $\Gamma$ in the bulk that ends on $\Sigma$, i.e.~that satisfies $\Sigma = \partial \Gamma$. To leading order in the bulk semiclassical limit, the entanglement entropy is then
\be
S_E = \frac{A_\Gamma}{4 G_N} \,,
\ee
where $A_\Gamma$ is the area of the bulk minimal surface, and $G_N$ is the bulk Newton's constant. The area will be divergent because the minimal surface reaches all the way to the asymptotic boundary. It is natural to cut off this divergence at a UV scale, where it provides the short distance ``boundary law'' of the entanglement entropy.

Holographic computations of the entanglement entropy appear rather similar to those of \emph{spatial} Wilson loops \cite{Maldacena:1998im, Rey:1998ik}. In $2+1$ field theory dimensions in particular, both quantities depend on a spatial curve $\Sigma_1$. However, there can be subtle differences. The entanglement entropy is given by the area of a minimal bulk spatial hypersurface in the Einstein frame metric. Areas of hypersurfaces in the Einstein frame have the important property that they are preserved under dimensional reduction. This allows the entanglement entropy to be computed without depending sensitively on the bulk UV completion. Spatial Wilson loops, in contrast, are computed by the bulk on-shell fundamental string worldsheet action in a full ten dimensional string theory background. This is consistent perhaps with the fact that the Wilson loop refers to a specific gauge-theoretic operator, whereas the entanglement entropy can be defined independently of the operator content of the quantum field 
theory.

The question of interest is then how $A_\Gamma$ scales with the linear dimension $R$ of the boundary region, in the limit of large $R$. There is an unfortunate linguistic clash here between the terms commonly employed for Wilson loops and entanglement entropy, with ``area law'' taken to mean different things. We will refer to a ``boundary law'' as a scaling
\be
A_\Gamma \sim R^{\, d-1} \,.
\ee
In contrast, a ``volume law'' will be
\be
A_\Gamma \sim R^{\, d} \,.
\ee
We have already noted that the entanglement entropy universally has a UV sensitive boundary law contribution due to short distance correlations across the surface $\Sigma$.

Geometries dual to confining theories typically have the property that the effective $(d+2)$-dimensional spacetime metric collapses in the far IR. In many circumstances this can be understood as an internal dimension capping off \cite{Witten:1998zw}. The collapse reduces the number of low energy degrees of freedom, typically leading to a mass gap.
In terms of the hypersurfaces needed to compute the entanglement entropy, the collapse of the geometry allows the hypersurfaces to terminate in the far IR, as their cross sectional areas vanish.

Consider the case where the hypersurface $\Sigma$ consists of two parallel infinite spatial hyperplanes separated by a distance $L$. We can think of this as the limit of a rectangle where all lengths except one have been taken large. With a confining gravity dual, there are then two candidate minimal surfaces $\Gamma$. One connects the two hyperplanes, while the other is disconnected, connecting each hyperplane separately to the far IR region where the bulk surfaces end. It has been found in several examples that at large separations $L$, the disconnected surfaces have a lower area and therefore determine the entanglement entropy \cite{Nishioka:2006gr, Klebanov:2007ws, Nishioka:2009un}. This leads to an entanglement entropy at large $L$ with a pure boundary form
\be\label{eq:cons}
S_E(L) \sim \frac{\text{Vol}\left(\Sigma \right)}{\e^{d-1}} + \text{Vol} \left(\Sigma \right) \cdot \text{const.} \,,
\ee
where $\e$ is a UV cutoff. In particular, the entanglement entropy is independent, at leading order, of the separation $L$. This process is illustrated in figure \ref{fig:confdeconf} below.

\begin{figure}[h]
\begin{center}
\includegraphics[height=150pt, width = \textwidth]{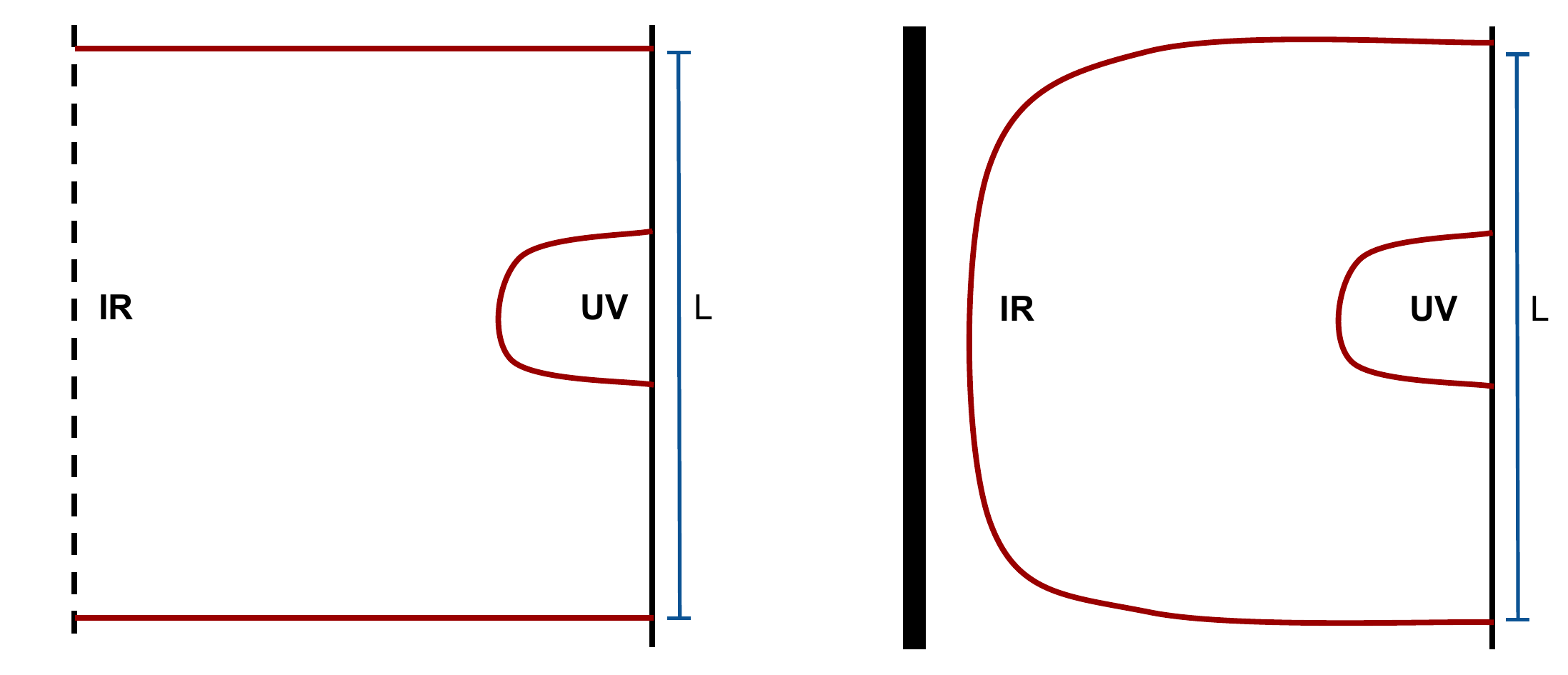}\caption{Left, minimal bulk hypersurfaces in a confining geometry. At large boundary separations $L$ the surface is disconnected and leads to a boundary law for the entanglement entropy. Right, minimal hypersurfaces in a finite temperature black hole geometry. At large separations $L$ the surface remains connected and leads to a volume law. In both cases, the surfaces for small boundary separations are connected. \label{fig:confdeconf}}
\end{center}
\end{figure}

Geometries dual to deconfined phases, in contrast, are typically characterized by an event horizon in the IR.
Horizons can be scale invariant, such as the AdS Poincar\'e horizon or the various extremal horizons discussed in detail below, or can be finite temperature Schwarzschild-like horizons. Zero temperature horizons are considered in the following paragraph. For finite temperature horizons, similarly to the case of confining geometries, there is an IR scale at which the geometry ends. However, the geometry close to the horizon is distinct: minimal spatial surfaces are unable to cross the horizon \cite{Ryu:2006bv, Nishioka:2009un, Hubeny:2012ry}. Instead, as the boundary hyperplanes we considered in the previous paragraph are taken far apart, the bulk surfaces droop down close to the horizon and then extend along the horizon. This leads to a volume law of the entanglement entropy at large separation
\be\label{eq:vol}
S_E(L) \sim \frac{\text{Vol}\left(\Sigma \right)}{\e^{d-1}} + L \cdot \text{Vol} \left(\Sigma \right) \cdot \text{const.} \,.
\ee
The second term here is the volume scaling term and is essentially just the thermal entropy of the system. This behavior is also illustrated in figure \ref{fig:confdeconf}.

For extremal horizons, while the spatial bulk surfaces are still unable to cross the horizon, the emergence of a scaling symmetry in the far IR typically results in a bulk surface that does not lead to a volume law. The IR correction to the UV sensitive boundary scaling is a term in the entanglement entropy that vanishes at large separations, scaling like $L^{\theta-(d-1)}$, where $\theta < d-1$ is the hyperscaling violation exponent \cite{Ogawa:2011bz, Huijse:2011ef}. We will derive this fact below. If $\theta = d - 1$ one finds a logarithmic enhancement to the boundary law \cite{Ogawa:2011bz, Huijse:2011ef}. One interesting exceptional case is the extremal Reissner-Nordstr\"om horizon, where a volume scaling term as in (\ref{eq:vol}) does appear due to the ground state entropy density and corresponding finite size horizon \cite{Swingle:2009wc}. Other exceptional cases are locally critical theories without a ground state entropy density, recently discussed in \cite{Hartnoll:2012wm}. In these cases, at large separation the bulk hypersurface 
appears to be disconnected, leading to an entanglement entropy of the form (\ref{eq:cons}), as for confining geometries.

\subsection{Charged sector: Cohesion and fractionalization}

The previous subsection reviewed how the entanglement entropy distinguishes between various confining and deconfined holographic phases. The essential point is that the bulk minimal surfaces ending on well separated parallel boundary hyperplanes probe the far IR spacetime geometry. These have qualitatively different behavior depending on whether the IR geometry collapses or ends at a horizon. We would like a refinement of the entanglement entropy that can determine whether there is flux emanating from behind a horizon, in addition to detecting the horizon itself.

\begin{figure}[h]
\begin{center}
\includegraphics[height=90pt, width = \textwidth]{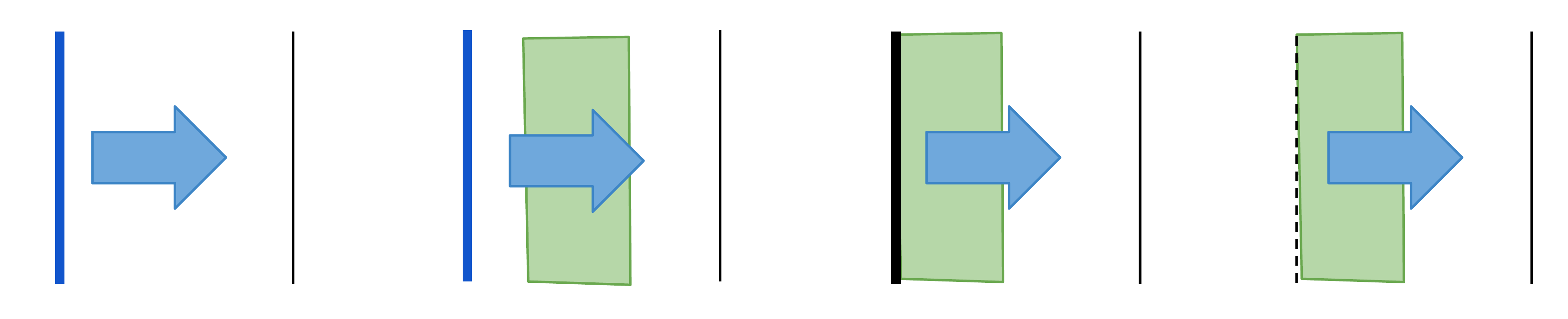}\caption{From left to right: (i) all flux emanates from horizon, (ii) a fraction of the flux emanates from a horizon, a fraction is from charged bulk fields, (iii) all flux from bulk fields outside a neutral horizon, (iv) all flux from bulk fields in a confining geometry. \label{fig:fraccohesive}}
\end{center}
\end{figure}

In figure \ref{fig:fraccohesive} below we show the various classes of bulk backgrounds that we would like to differentiate. We are primarily interested in zero temperature solutions. There are at least four possibilities. If there is a horizon in the spacetime, then all or some of the electric flux may emanate from behind the horizon (we will include certain null singularities in our notion of horizons). These are the {\bf fractionalized} and {\bf partially fractionalized} cases. If, in contrast, all flux originates from charged matter in the spacetime, then there may either be a neutral horizon in the far IR, or the geometry may confine. These are {\bf deconfined cohesive} and {\bf confined cohesive} phases, respectively.
One may find in addition spacetimes in which the charge emanates from a singularity, but with a geometry that is confining according to the usual criteria \cite{Gubser:2000nd}. Known examples have apparent pathologies such as hyperscaling exponent $\theta > d-1$. For this reason we have not included a picture corresponding to confined but fractionalized phases. It is of interest to see if such phases can exist.

All of the possibilities shown in figure \ref{fig:fraccohesive} have been constructed. Fully fractionalized phases include extremal black holes with and without dilaton fields \cite{Chamblin:1999tk, Cvetic:1999ne,Taylor:2008tg, Goldstein:2009cv, Charmousis:2010zz, Gouteraux:2011ce}. Partially fractionalized backgrounds were constructed in \cite{Hartnoll:2011pp, D'Hoker:2012ej}. Backgrounds with deconfined neutral horizons and cohesive charge carriers include electron stars \cite{Hartnoll:2009ns, Hartnoll:2010gu, Allais:2012ye}, the ground states of holographic superconductors \cite{Gubser:2009cg, Horowitz:2009ij} and the topologically charged spacetimes of \cite{D'Hoker:2012ej}. Confined cohesive phases include electron stars and holographic superconductors in confining spacetimes \cite{Nishioka:2009zj, Horowitz:2010jq, Arsiwalla:2010bt, Bhattacharya:2012we}.

The entanglement entropy alone is unable to distinguish between charged and neutral horizons. This is clear because it is possible to have the same near-horizon geometry with and without flux emanating from behind the horizon (compare for instance \cite{Goldstein:2009cv} and \cite{Hartnoll:2010gu}). The area $A_\Gamma$ of a bulk spatial
hypersurface $\Gamma$ is the simplest invariant one can associate to the surface. However, given a bulk theory with a Maxwell field, a second equally simple invariant is the conserved electric flux $\Phi_\Gamma$ through the hypersurface. The origin of the electric flux in the bulk is precisely the question we are trying to address. Therefore a flux-sensitive observable is likely to be of use. This motivates consideration of the following `deformed' holographic entanglement entropy
\be\label{eq:qS}
S_E^{\, \g} = \frac{A_\Gamma}{4 G_N} + \g \, \Phi_\Gamma \,.
\ee
The idea is that we will again associate a bulk spatial hypersurface $\Gamma$ to a boundary spatial hypersurface $\Sigma$  such that $\pa \Gamma = \Sigma$. The deformed entropy (\ref{eq:qS}) is to be evaluated on this bulk surface $\Gamma$. In the remainder we will discuss exactly which bulk hypersurface $\Gamma$ should be used in the expression for the deformed entropy. We will consider two scenarios. In the first, we take the surface $\Gamma$ to be the same Ryu-Takayanagi minimal surface as is used for the entanglement entropy. In this case, the flux $\Phi_\Gamma$ can be thought of as additional data associated to the entanglement entropy. In the second scenario, we extremize the deformed entropy itself. Here $\Phi_\Gamma$ is to be thought of as an electric dipole moment that has been induced on a D-brane worldvolume theory. Such polarized D-branes would presumably be dual to a class of surface operators.

\section{A hypersurface order parameter}

For any bulk spatial hypersurface $\G$, the flux $\Phi_\Gamma$ is easily evaluated in fully fractionalized cases. The absence of explicit charged sources, combined with Gauss's law and assuming for the moment that the surface does not fall through the horizon (we recalled above that minimal spatial hypersurfaces indeed cannot), leads to the conclusion that
\be\label{eq:phifrac}
\Phi_\Gamma = \rho \, L \cdot \text{Vol} \left(\Sigma \right) \,.
\ee
Thus $\Phi_\Gamma$ obeys a volume law in these cases. Here $\rho$ is the charge density of the boundary field theory and we have again taken the asymptotic boundary $\Sigma$ of $\Gamma$ to be two very large parallel spatial hyperplanes separated by a distance $L$. Recall that the field theory charge density is given by the electric flux at the asymptotic boundary \cite{Hartnoll:2011fn}. The result (\ref{eq:phifrac}) is\ proven in figure \ref{fig:fluxgauss} below.

\begin{figure}[h]
\begin{center}
\includegraphics[height=190pt]{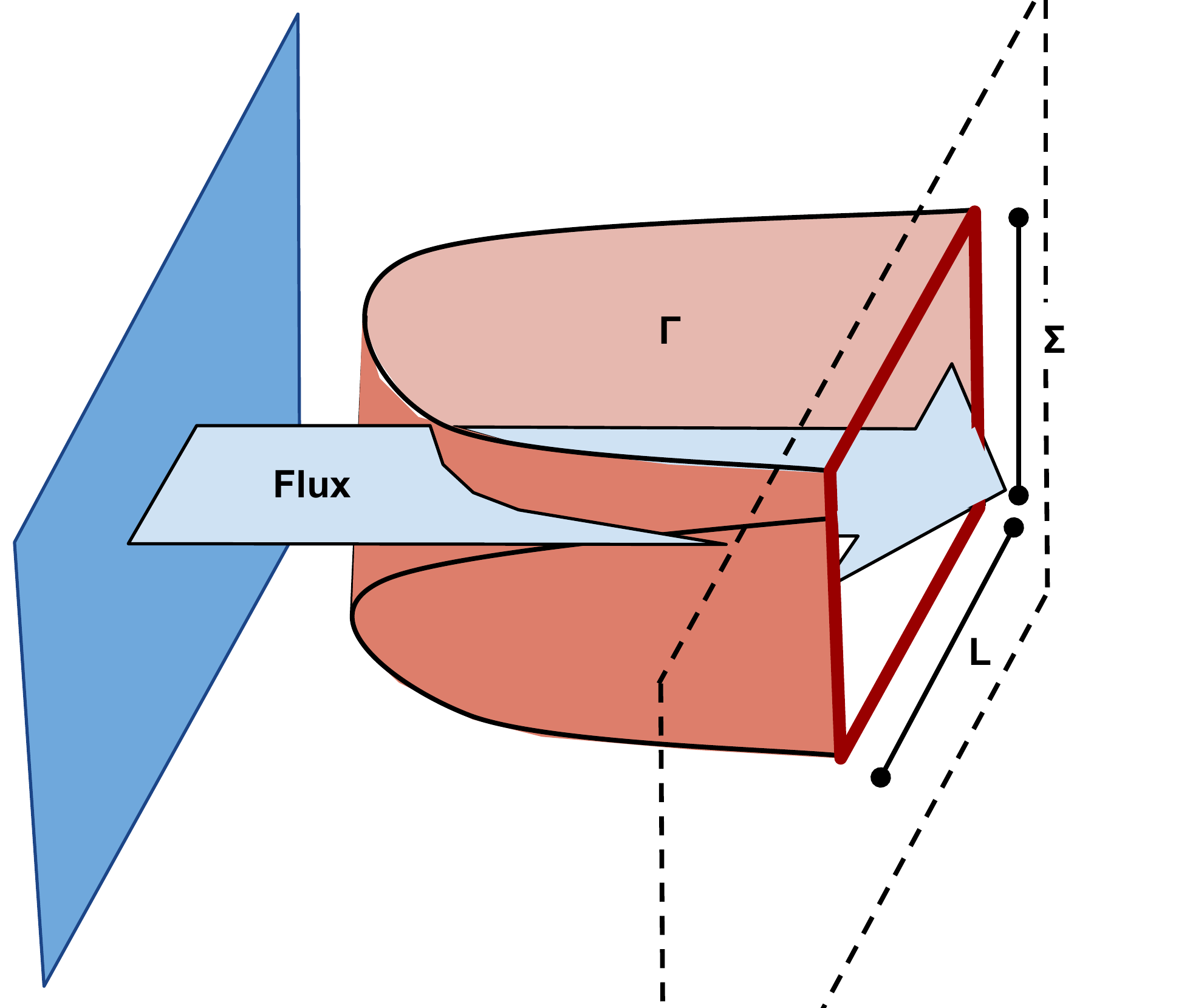}\caption{By Gauss's law, in the absence of bulk charges, the electric flux through the surface $\Gamma$ is equal to the flux  at the asymptotic boundary in the region bounded by the surface $\Sigma = \pa \Gamma$. The total asymptotic flux is $\rho \, L \cdot \text{Vol} \left(\Sigma \right)$. \label{fig:fluxgauss}}
\end{center}
\end{figure}

By $\Phi_\Gamma$ we will always refer to the conserved flux. In the presence of bulk dilaton fields, this will not be simply the integral of the Maxwell field strength. Gauss's law, using if necessary a modified flux due to dilatons \cite{Hartnoll:2011pp, Hashimoto:2012ti}, plays a key role in holographic derivations of the Luttinger theorem \cite{Iqbal:2011bf, Sachdev:2011ze}. It seems natural that it should also feature prominently in our discussion.

Another situation that is easy to describe is when there are disconnected bulk surfaces that fall straight into the IR and terminate there, such as the minimal surfaces in confining geometries that we reviewed above. In this case, independently of whether the flux originates from a horizon or from charged bulk matter, there is no flux through the bulk hypersurface and so
\be\label{eq:phiconf}
\Phi_\Gamma = 0 \,.
\ee
Here we are making the assumption that the bulk flux is purely radial and that the bulk surfaces fall straight into the IR without bending and picking up some flux.

The cases that remain to be considered are those in which at least part of the flux is sourced by charged matter in the bulk and where the hypersurface $\Gamma$ is not of the disconnected type described in the previous paragraph. These cases are more sensitive to the precise form of the surface involved. In the following two subsections we will consider two distinct natural choices for the bulk hypersurface. As we discuss in the final section, these correspond to two distinct dual field theoretic interpretations of the flux.

\subsection{Electric flux through minimal hypersurfaces in the bulk}

Minimal spatial hypersurfaces in the bulk determine the entanglement entropy of boundary regions according to the Ryu-Takayanagi proposal reviewed above. The simplest choice of hypersurface $\G$ on which to evaluate the deformed entropy would thus seem to be these minimal hypersurfaces. From this perspective, the deformed entropy is not really being taken seriously as a new observable; rather we are associating a new quantity, the flux through the bulk hypersurface, to the data that arises in computing the entanglement entropy holographically. 

We have already computed the flux through minimal hypersurfaces in fully fractionalized (\ref{eq:phifrac}) and in confining phases (\ref{eq:phiconf}). To discuss partially fractionalized and deconfined cohesive phases, we need to find the explicit form of the hypersurface in the far IR. This depends of course on the spatial IR geometry of the bulk spacetime. We will consider a large class of zero temperature scaling geometries that have arisen in recent holographic studies as the near horizon geometries of extremal black holes, both with \cite{Hartnoll:2009ns, Hartnoll:2010gu, Hartnoll:2011pp, D'Hoker:2012ej} and without
\cite{Taylor:2008tg, Goldstein:2009cv, Charmousis:2010zz}
charged bulk matter. The IR spatial geometry takes the form
\be\label{eq:space}
ds^2_\text{spatial} = \frac{dr^2}{r^{2(d-2\theta)/(d-\theta)}} + \frac{dx^2_d}{r^2} \,.
\ee
Here the IR limit is understood to mean $r \to \infty$. The boundary spatial dimension is $d$ and $\theta$ is the hyperscaling violation exponent \cite{Huijse:2011ef, Gouteraux:2011ce}. The dynamical critical exponent $z$ will not appear in our discussion.

We wish to consider hypersurfaces that are extended along $d-1$ of the boundary spatial dimensions, forming a large rectangle with area $\text{Vol} \left(\Sigma \right)$, and that form a curve $x(r)$ in the remaining dimension. In that case the area of the hypersurface is given by
\be\label{eq:area}
A = \text{Vol} \left(\Sigma \right) \int \frac{dr}{r^d} \sqrt{r^{2\theta/(d-\theta)} + \dot x^2} \,.
\ee
A minimal hypersurface $\G$ is then easily found to solve
\be
\dot x^2 = \frac{r^{2\theta/(d-\theta)}}{\left(r_o/r\right)^{2d} - 1} \,.
\ee
Here $r_o$ is the maximal radius reached by the hypersurface.

For the case of large boundary separation $L$ in the $x$ direction, we can relate $L$ to the depth $r_o$ that the hypersurface reaches
\be\label{eq:L}
L = \int dr \dot x \sim \int^{r_o} dr \frac{r^{\theta/(d-\theta)}}{\sqrt{\left(r_o/r\right)^{2d} - 1}} \sim r_o^{d/(d-\theta)} \,.
\ee
In particular, because $\theta < d-1$ \cite{Huijse:2011ef}, $L$ will diverge as $r_o \to \infty$. This shows us that picking out the leading IR contribution, i.e. the one in  the range $r \sim r_o$, is self-consistent. The region $0 \leq r \ll r_o$ of the integral (\ref{eq:L}) will give a contribution to $L$ that does not diverge as $r_o \to \infty$. This is good because the IR contribution is universal in that it only depends on the scaling geometry (\ref{eq:space}), while the remaining contribution is sensitive to the whole bulk spacetime. The case $\theta \to - \infty$ is subtle and needs to be considered separately \cite{Hartnoll:2012wm}.

We may now extract the universal IR contribution to the area and flux through the hypersurface.
From (\ref{eq:area}), the area is
\be\label{eq:Ascale}
\frac{A_\Gamma}{\text{Vol} \left(\Sigma \right)} \sim \int^{r_o} \frac{dr}{r^d} \frac{r^{\theta/(d-\theta)} \cdot (r_o/r)^d}{\sqrt{\left(r_o/r\right)^{2d} - 1}} \sim r_o^{d(1-d+\theta)/(d-\theta)} \sim L^{\theta - (d-1)}  \,.
\ee
While this term necessarily vanishes at large $L$, again because $\theta < d-1$, it represents the leading interesting large $L$ dependence of the area of the hypersurface. The leading contribution from the remainder of the spacetime will just be a constant that is independent of $L$. This constant will include the UV divergent boundary law term. The picture that one has in mind is that, at large separations $L$, the minimal surface falls straight into the IR region before it has the chance to significantly bend and close off. As we have discussed above, the behavior (\ref{eq:Ascale}) of the hypersurface area is only sensitive to the spatial geometry (\ref{eq:space}) and is not sensitive to the origin of the electric flux.

To obtain the flux through the bulk minimal surface we need to know the distribution of bulk charges. In all known cohesive deconfined cases, where the charge extends all the way into the IR, the charge density is constant in the scaling regime where the geometry has the form (\ref{eq:space}) \cite{Hartnoll:2009ns, Hartnoll:2010gu, Hartnoll:2011pp, Gubser:2009cg, Horowitz:2009ij, D'Hoker:2012ej}. This of course befits a scalar quantity in a scaling solution. In known partially fractionalized solutions, the charge ends a certain radius and does not extend all the way into the IR \cite{Hartnoll:2011pp}. We consider these two cases separately.

In the presence of a constant charge density $\sigma$ in the IR geometry, Gauss's law for the (generalized if necessary) electric field $E(r)$ takes the form
\be
\frac{d}{dr} E = \frac{\sqrt{g_{rr}}}{r^d} \sigma \,.
\ee
Therefore in a cohesive deconfined phase, with a neutral horizon and spatial geometry (\ref{eq:space}), the electric field is
\be\label{eq:EE}
E = \int_r^\infty \frac{\sqrt{g_{rr}}}{r^d} \sigma dr \sim \sigma r^{(\theta d + \theta - d^2)/(d-\theta)} \,.
\ee
The IR contribution to the flux through the minimal surface is then
\be
\frac{\Phi_\Gamma}{\text{Vol} \left(\Sigma \right)} = \int E dx = \int^{r_o} \frac{\sigma \, \dot x \, dr}{r^{(d^2 - \theta d - \theta)/(d-\theta)}}
\sim \sigma \, r_o^{[d(1-d+\theta) + \theta]/(d-\theta)} \sim \sigma \, L^{\theta+\theta/d - (d-1)} \,.  
\ee
We see that, unlike the area, the IR contribution to the flux can grow with $L$. Specifically, it grows if $d(d-1)/(d+1) < \theta$. However, given that $\theta < d-1$, it is always less than a volume law.

In partially fractionalized cases the charge density does not reach the far IR. From our observation in equation (\ref{eq:L}), that $L$ grows with $r_o$, it follows that the leading IR contribution to the flux will be insensitive to the presence of bulk charges. This is because the surface has `fallen through' the region where the charges lie. Therefore $E_\text{IR}$ is constant and the flux through the minimal surface is
\be
\frac{\Phi_\Gamma}{\text{Vol} \left(\Sigma \right)} = \int E dx \sim E_\text{IR} L \sim (\rho - \rho_\text{coh.}) L \,.
\ee
For the final step, we used the fact that Gauss's law implies that $\rho = E_\text{UV} = E_\text{IR} + \rho_\text{coh.}$, where $\rho_\text{coh.}$ is the total boundary field theory charge density that is accounted for as conventional gauge-invariant `cohesive' charge. This is the charge density that contributes to the Luttinger count, in the case of fermionic matter. It follows that partially fractionalized phases exhibit the same volume law scaling for the flux $\Phi_\Gamma$ as fractionalized phases, but with a coefficient that is less than the total charge density.

The results we have just discussed are summarized in table \ref{default} below.
\begin{table}[h]
\begin{center}
\vspace{0.6cm}
\begin{tabular}{|c|c|c|}
\hline
Holographic phase &  IR contribution to area $A_\Gamma$ & IR contribution to flux $\Phi_\Gamma$ \\
\hline
Fully fractionalized & $L^{\theta - (d-1)}  \cdot \text{Vol} \left(\Sigma \right) \,$ or $\, L \cdot \text{Vol} \left(\Sigma \right)$ & $\rho \, L \cdot \text{Vol} \left(\Sigma \right)$ \\
Partially fractionalized & $L^{\theta - (d-1)}  \cdot \text{Vol} \left(\Sigma \right)$ & $\left( \rho - \rho_\text{coh.} \right) \, L \cdot \text{Vol} \left(\Sigma \right)$ \\
Deconfined, cohesive & $ L^{\theta - (d-1)}  \cdot \text{Vol} \left(\Sigma \right)$ & $L^{\theta + \theta/d - (d-1)}  \cdot \text{Vol} \left(\Sigma \right)$\\
Confined, cohesive & $\text{Vol} \left(\Sigma \right)$ & 0 \\
\hline
\end{tabular}
\caption{The IR contribution to the area and electric flux through a minimal spatial hypersurface $\G$ in the bulk, in different holographic phases. The boundary of the hypersurface $\partial \G = \Sigma$ is two spatial hyperplanes separated by a large distance $L$. \label{default}}
\end{center}
\end{table}
We see that the electric flux through minimal surfaces in the bulk is able to distinguish the four phases of interest.
Fractionalized and partially fractionalized phases exhibit a volume law for the flux term, with a coefficient that detects the extent of fractionalization. Deconfined but cohesive phases have a flux given by a universal $L$ dependence that is always less than a volume law. Confined and cohesive phases have no flux. 

In addition to the IR contributions shown in table \ref{default}, there will be a contribution from the remainder of the surface. An exception to this statement is for the confined cohesive case, in which $\Phi_\Gamma = 0$ identically. The nonuniversal contribution will be proportional to $\text{Vol} \left(\Sigma \right)$ but independent of $L$ to leading order at large separations. This provides a boundary law contribution. Thus we can also summarize the results as follows:
\bea\label{eq:fract}
\text{Fractionalization} \quad & \Rightarrow & \quad \text{$\Phi_\Gamma \sim$  volume law} \,, \\ \label{eq:deconf}
\text{Deconfined cohesive} \quad & \Rightarrow & \quad \text{boundary law $\leq \Phi_\Gamma <$ volume law} \,, \\ \label{eq:conf}
\text{Confined cohesive} \quad & \Rightarrow & \quad \Phi_\Gamma = 0 \,.
\eea
We emphasize that the volume law for the flux here is independent of $\theta$ and $z$ in the near horizon geometry, and in particular is not related to the volume law for the entanglement entropy of extremal Reissner-Nordst\"om black holes.

\subsection{Minimizing the deformed entropy}

Taking the deformed holographic entanglement entropy more seriously, we can select the bulk hypersurface $\Gamma$ by minimizing the deformed entropy itself. Evaluating the deformed entropy \eqref{eq:qS} on the scaling geometry background (\ref{eq:space}), and using the result (\ref{eq:EE}) for the electric field in terms of the background charge density $\sigma$, we obtain the functional
\be
  S^\g_E = \frac{A_\Gamma}{4 G_N} + \g \, \Phi_\Gamma = \text{Vol} \left(\Sigma \right) \int dr\ \left( \frac1{4G_N r^d}\sqrt{r^{2\theta/(d-\theta)} + \dot x^2} + \g \frac{\sigma \, \dot x}{r^{(d^2 - \theta d - \theta )/(d-\theta)}}\right)\, .
\ee
Minimizing the deformed entropy thus reduces to studying the dynamics given by the Lagrangian
\be\label{eq:llag}
  \Lag = \frac1{r^d} \left(\frac1{4G_N}\sqrt{r^{2\theta/(d-\theta)} + \dot x^2} + \g \sigma \, \dot x \, r^{\theta/(d-\theta)} \right) \, .
\ee
While we will continue to use the language of entropy, and the associated $1/(4 G_N)$ prefactor of the area, it is presumably most natural to view this Lagrangian as describing the worldvolume dynamics of a Euclidean brane that is electrically neutral but carries an inherent electric dipole moment. The flux through the surface $\Phi_\Gamma \sim \int_\Gamma \star F$ precisely describes the coupling of a dipole moment on the surface to the external electric field.

The quantity $C \equiv \del \Lag/\del\dot x$ is constant because the Lagrangian (\ref{eq:llag}) does not depend on $x(r)$. Knowing this, it is easy to determine that the profile of the hypersurface satisfies
\be\label{eq:xdotsq}
  \dot x^2 = r^{2\theta/(d - \theta)} \frac{D^2(r)}{(1/4G_N)^2 - D^2(r)}\, ; \quad D(r) \equiv C r^d - \g \sigma\, r^{\theta/(d - \theta)}\, .
\ee
The maximal radius $r_o$ reached by the hypersurface is given by $\dot x^2(r_o) = \infty$ or, equivalently,
\be\label{eq:endd}
  \big(4G_N D(r_o)\big)^2 = 1\,.
\ee
For $\g \sigma = 0$, this condition reduces to $C = \pm 1/4G_N r_o^d$, and we recover the equation of motion found in the previous subsection. Thus, in particular, for fully fractionalized geometries, the computation and results are the same as before. The sign of $C$ is irrelevant in this case. However, for $\g \sigma \neq 0$, the parameter space needs to be carefully studied. Not all values of $\g \sigma$ need yield a single hypersurface that starts at the boundary $r = 0$ and ends at $r = r_o$; depending on the sign of $C$, it is possible to have none or several solutions. We will explore such interesting dynamical features of the Lagrangian (\ref{eq:llag}) in the following subsection.

In this subsection we wish to determine the IR contribution to the area $A_\Gamma$ and flux $\Phi_\Gamma$ associated to the hypersurface $\Gamma$ that minimizes the deformed entropy. We focus on the IR contribution, in the cases where this exists, by restricting all relevant integrals to the region $r_o(1 - \e) \leq r \leq r_o$, with the depth $r_o$ taken to be large. At small $\e$, we may expand $D(r) = D(r_o) - D'(r_o) (r_o - r)$, with $D(r_o) = 1/4G_N$ and
\be
  D'(r_o) \sim \begin{cases}
		\frac1{r_o}, & \theta \leq 0 \, ,\\
		\frac1{r_o}r_o^{\theta/(d - \theta)}, & \theta \geq 0 \, .
               \end{cases}
\ee
The different cases arise due to different terms balancing at leading order in the equation that determines the endpoint.
To leading order in $\e$, the boundary separation associated to depth $r_o$ is
\be
  L \sim \int_{r_o(1 - \e)}^{r_o} \!\!\!\!dr\, \dot x \sim r_o^{\theta/(d - \theta)}\sqrt{\frac{r_o D(r_o)}{D'(r_o)}} \sim
  \begin{cases}
    r_o^{d/(d - \theta)}, & \theta \leq 0\, , \\
    r_o^{(2d - \theta)/2(d - \theta)}, & \theta \geq 0 \, .
  \end{cases}
\ee
In both cases $L$ diverges as $r_o \to \infty$, justifying the isolation of the near boundary contribution to capture the large separation behavior. At this point, we are only considering values of $G_N$ and $\g \sigma$ such that a solution to $D(r_o) = 1/4G_N$ exists at large $r_o$ for some choice of constant $C$.

We can now evaluate the two terms in the action on these solutions. The flux is
\be
  \Phi_\Gamma \sim \int_{r_o(1 - \e)}^{r_o} \!\frac{dr}{r^d} \dot x\,  r^{\theta/(d-\theta)} \sim
  \begin{cases}
    r_o^{(d + \theta)/(d - \theta) - d} \sim L^{1 - d + \theta + \theta/d}, & \theta \leq 0 \, , \\
    r_o^{(2d + \theta)/2(d - \theta) - d} \sim L^{(\theta + 2d(1- d + \theta))/(2d - \theta)}, & \theta \geq 0 \, ,
  \end{cases}
\ee
and the area is
\be
  A_\Gamma \sim \int_{r_o(1 - \e)}^{r_o} \!\frac{dr}{r^d} \sqrt{\dot x^2 +  r^{2\theta/(d-\theta)}} \sim
  \begin{cases}
    r_o^{d/(d - \theta) - d} \sim L^{1 - d + \theta}, & \theta \leq 0 \, , \\
    r_o^{\theta/2(d - \theta) - d + 1} \sim L^{(-\theta + 2d(1- d + \theta))/(2d - \theta)}, & \theta \geq 0 \, .
  \end{cases}
\ee
We see that, with $\theta < d-1$, the area term dominates the on shell action for $\theta < 0$, while the flux term dominates for $\theta > 0$. Both terms contribute equally when $\theta = 0$. Furthermore, both terms typically go to zero with large $L$. There is a window of positive values of $\theta$ just below $\theta = d - 1$ where the action becomes large with large $L$. The power of $L$ in these cases is always less than one. Thus, as we concluded in the previous section for the flux through minimal surfaces (eq.~\eqref{eq:deconf}), allowing for the contribution of a non-universal boundary scaling term we have
\be
\text{Deconfined cohesive} \quad  \Rightarrow  \quad \text{boundary law $\leq S^\gamma_E <$ volume law} \,.
\ee

In fractionalized phases, the absence of bulk charge combined with Gauss's law implies that the flux term in the deformed entropy is a total derivative that does not affect the dynamics. The surface $\Gamma$ will therefore simply minimize the area. The same conclusion holds for the far IR of partially fractionalized backgrounds, where there is again no bulk charge. Our discussion in previous sections thus goes through unchanged and we conclude, as in eq.~\eqref{eq:fract}, that the deformed entropy has a volume law scaling in these cases:
\be
\text{Fractionalization} \quad  \Rightarrow  \quad S^\gamma_E \sim \text{volume law} \,.
\ee

Finally, we consider the deformed entropy in confined cohesive phases. Let us consider the specific model of a fluid of charged fermions in an AdS soliton-like geometry, in which an internal $S^1$ caps off in the IR \cite{Bhattacharya:2012we}. The spatial bulk geometry in the far IR is then
\be
  ds^2 = \frac{\alpha \, dr^2}{1 - r} + (1 - r)d\theta^2 + \frac{dx_{d}^2}{r^2} \, .
\ee
We have rescaled the radial coordinate so that the $\theta$ circle collapses at $r = 1$. The deformed entropy is now given by
\be\label{eq:sec}
  S_E^\gamma  = 2\pi \text{Vol} \left(\Sigma \right) \int dr\ \left( \frac1{4G_N r^{d}}\sqrt{\alpha r^2 + (1 - r)\dot x^2} + \g \frac{\sigma\sqrt\alpha}{r^{d-1}}\dot x\right).
\ee
The constant of motion is 
\be
  C
  = \frac1{r^{d - 1}}\left(\g\s\sqrt\a + \frac1{4G_N} \frac{(1-r)\dot x/r^2}{\sqrt{\alpha + (1-r)\dot x^2/r^2}} \right),
\ee
and the Euler-Lagrange equation can be compactly written as
\be
  \dot x^2 = \alpha \frac{r^2}{1-r} \frac{1}{K^2(r) - 1};\quad K(r) \equiv \frac1{4G_N} \frac{\sqrt{1-r}}{r} \frac1{Cr^{d - 1} - \g\s\sqrt\a} \, .
\ee
As $r$ approaches the IR cutoff at unity, the denominator $K^2(r) - 1$ becomes negative \emph{unless} the constant $C$ takes on the precise value $C = \g\s\sqrt\a$. Only the hypersurface with this constant of motion probes the deep IR; all others terminate before reaching the confinement scale. This unique IR-probing hypersurface has $K^2(r)$ diverging as $1/(1-r)$ in the far IR, and its profile in this region satisfies
\be\label{eq:xd}
  \dot x = \pm 4(d - 1)\a\g\s G_N \,.
\ee

If this hypersurface is obtained by minimizing the undeformed entanglement entropy, we set $\g = 0$ and find that $\dot x = 0$, namely the hypersurface falls straight into the IR and terminates there. This is in keeping with our previous comments concerning minimal surfaces in confining geometries. It is interesting that the deformation of the entropy makes the hypersurface profile become linear with $r$ near the IR cutoff. This means that the entire bulk hypersurface gets ``skewed'' to one side in order to extremize the amount of flux it can get. Depending on the sign of $\g \sigma$, it is clear that one sign in (\ref{eq:xd}) will have lower deformed entropy (\ref{eq:sec}). In both cases, the minimizing surface at large separation has two disconnected components that fall into the far IR. Because the surfaces bend and pick up some flux in this case, we obtain a boundary law contribution to the flux as well as to the area of the surface.
\be
\text{Confined cohesive} \quad  \Rightarrow  \quad S^\gamma_E \sim \text{boundary law} \,.
\ee
This is different to the vanishing flux we found through minimal hypersurfaces in confining geometries (eq.~\eqref{eq:conf}). The deformed entropy can still distinguish confined cohesive and deconfined cohesive phases, however. In the confined cohesive case, there is no dependence at all on $L$ at large separations, while the boundary law in deconfined cohesive phases will have subleading $L$ dependence.

\subsection{Dynamics of dipole surfaces in charged media}

Before moving on to discuss possible field theory interpretations of the above results, we will explore in more detail the dynamics of hypersurfaces $\G$ that minimize the deformed entropy. This will be the dynamics of surfaces with a tension and a dipole moment in scale-invariant charged media.

\begin{figure}[!h]
 \centering
 \mbox{\includegraphics[width = 0.48\textwidth, height = 0.35\textwidth]{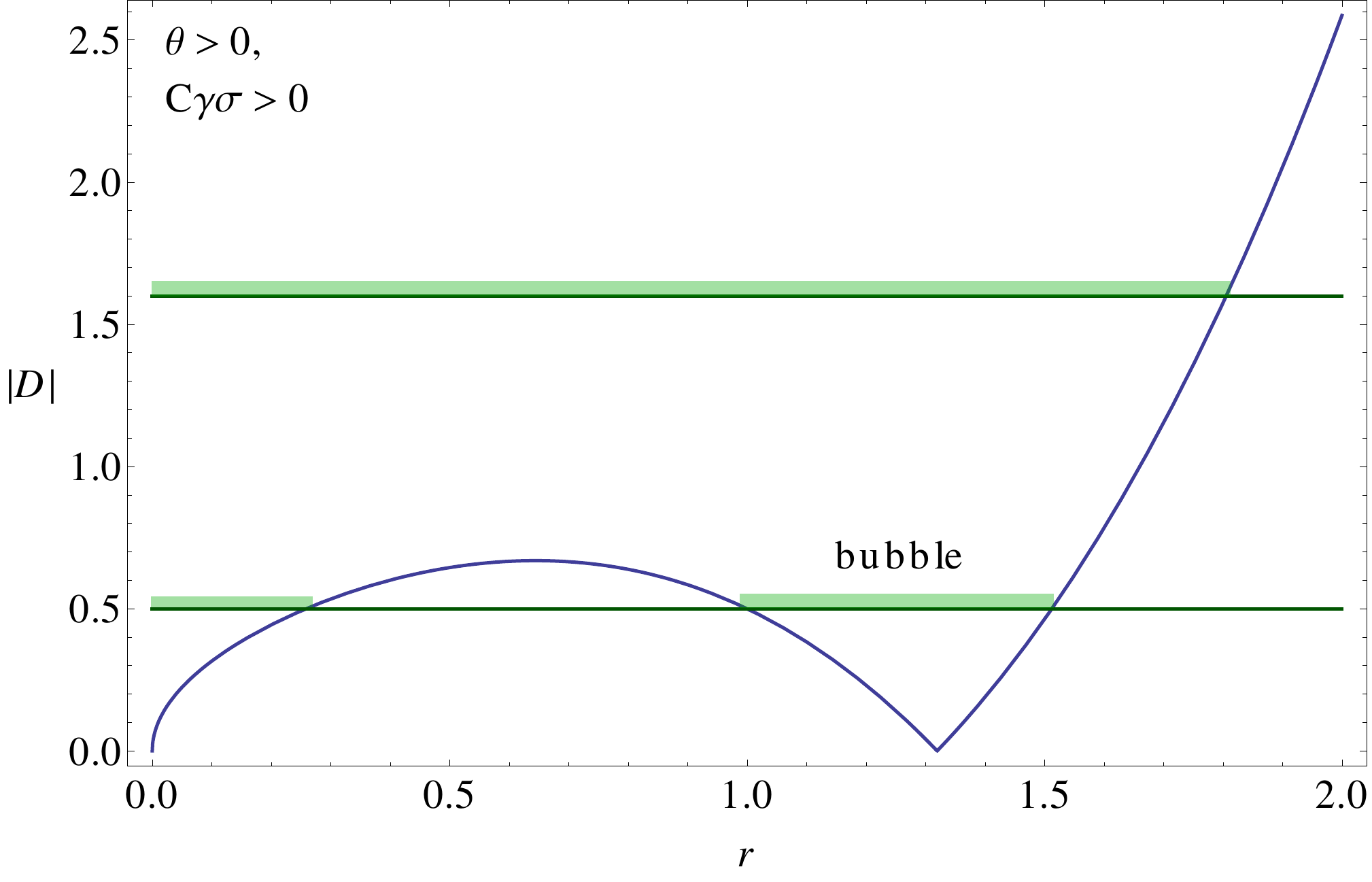}\hspace{0.02\textwidth} \includegraphics[width = 0.48\textwidth, height = 0.35\textwidth]{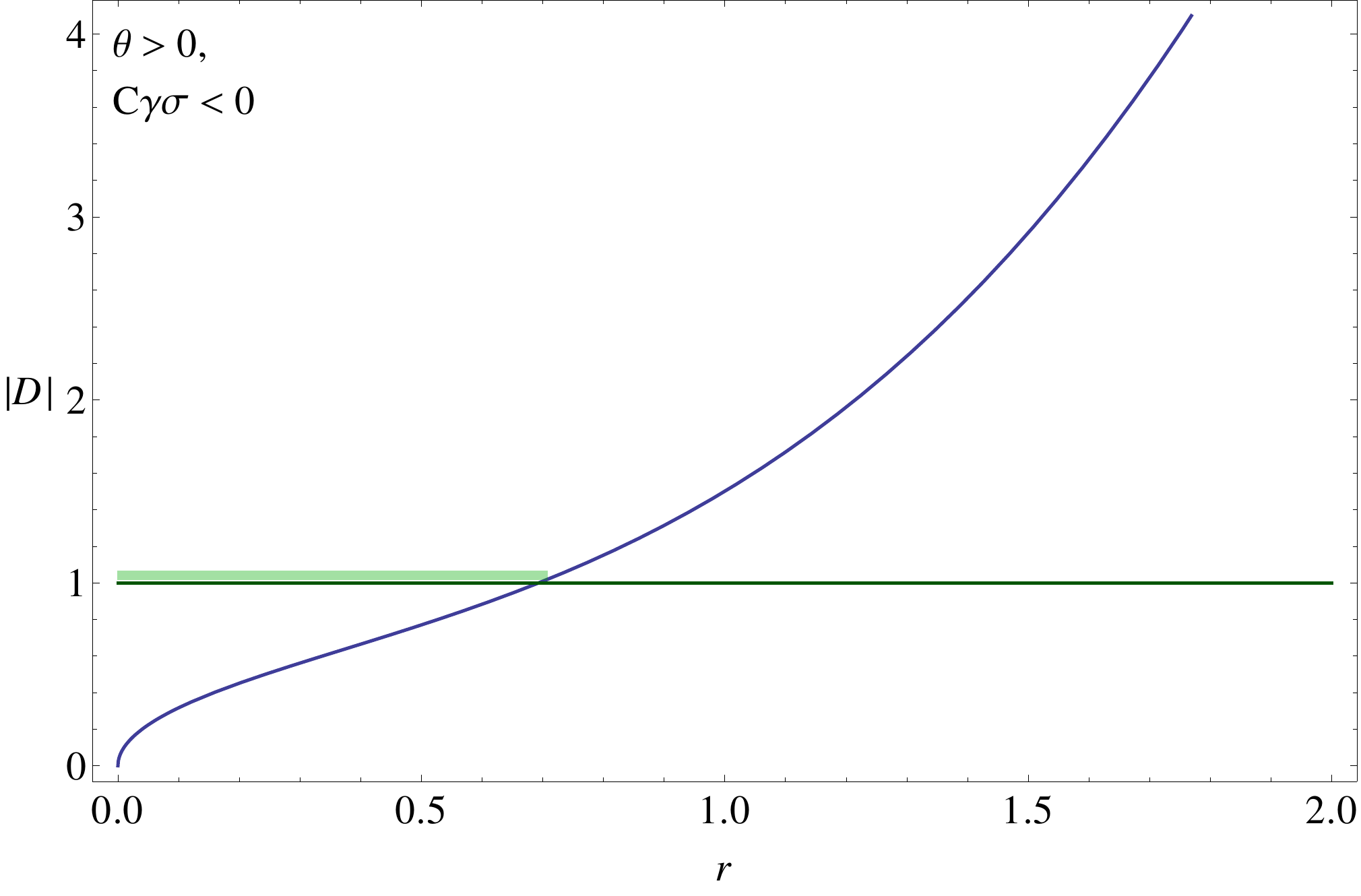}}
 \mbox{\includegraphics[width = 0.48\textwidth, height = 0.35\textwidth]{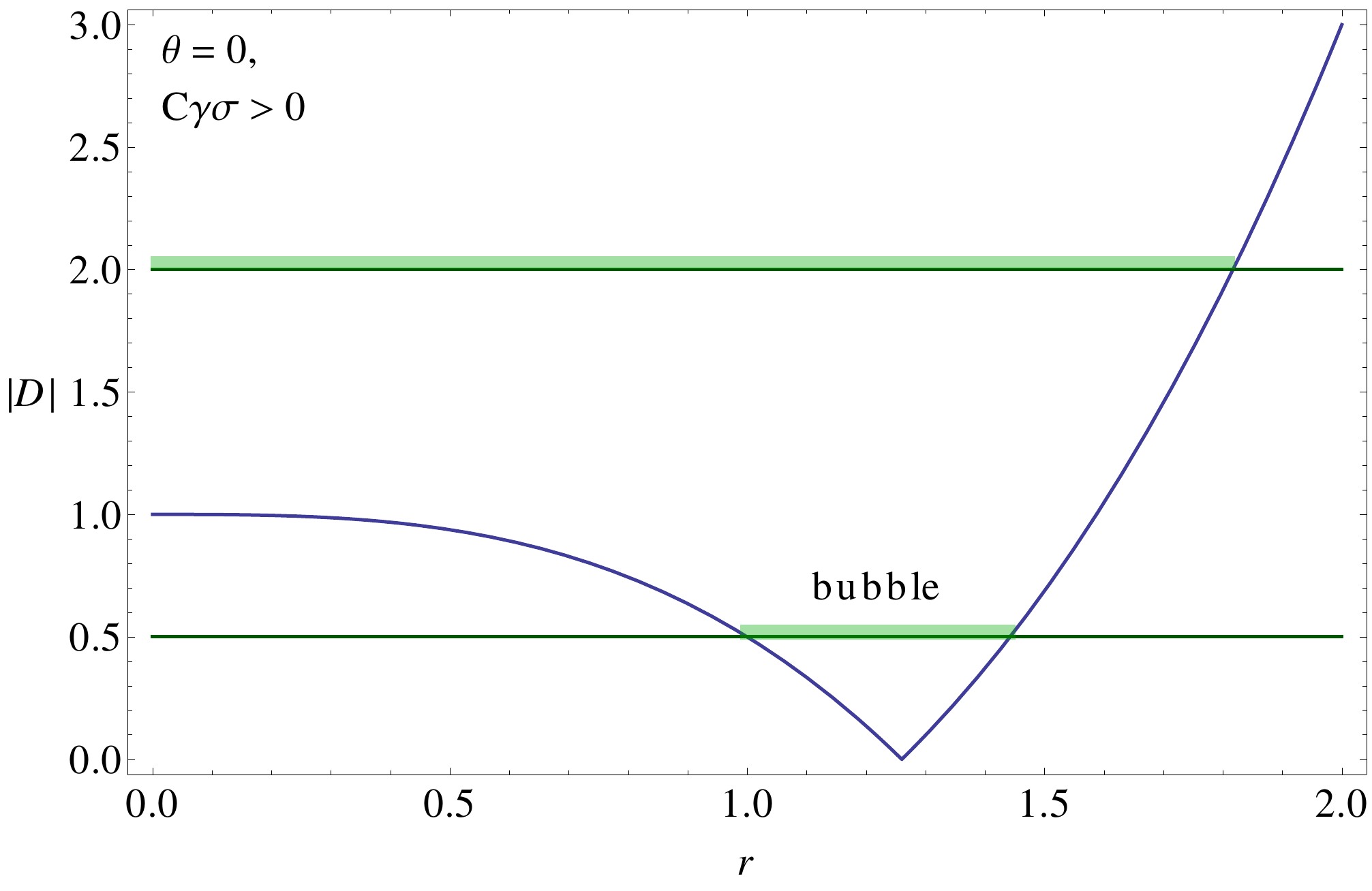}\hspace{0.02\textwidth} \includegraphics[width = 0.48\textwidth, height = 0.35\textwidth]{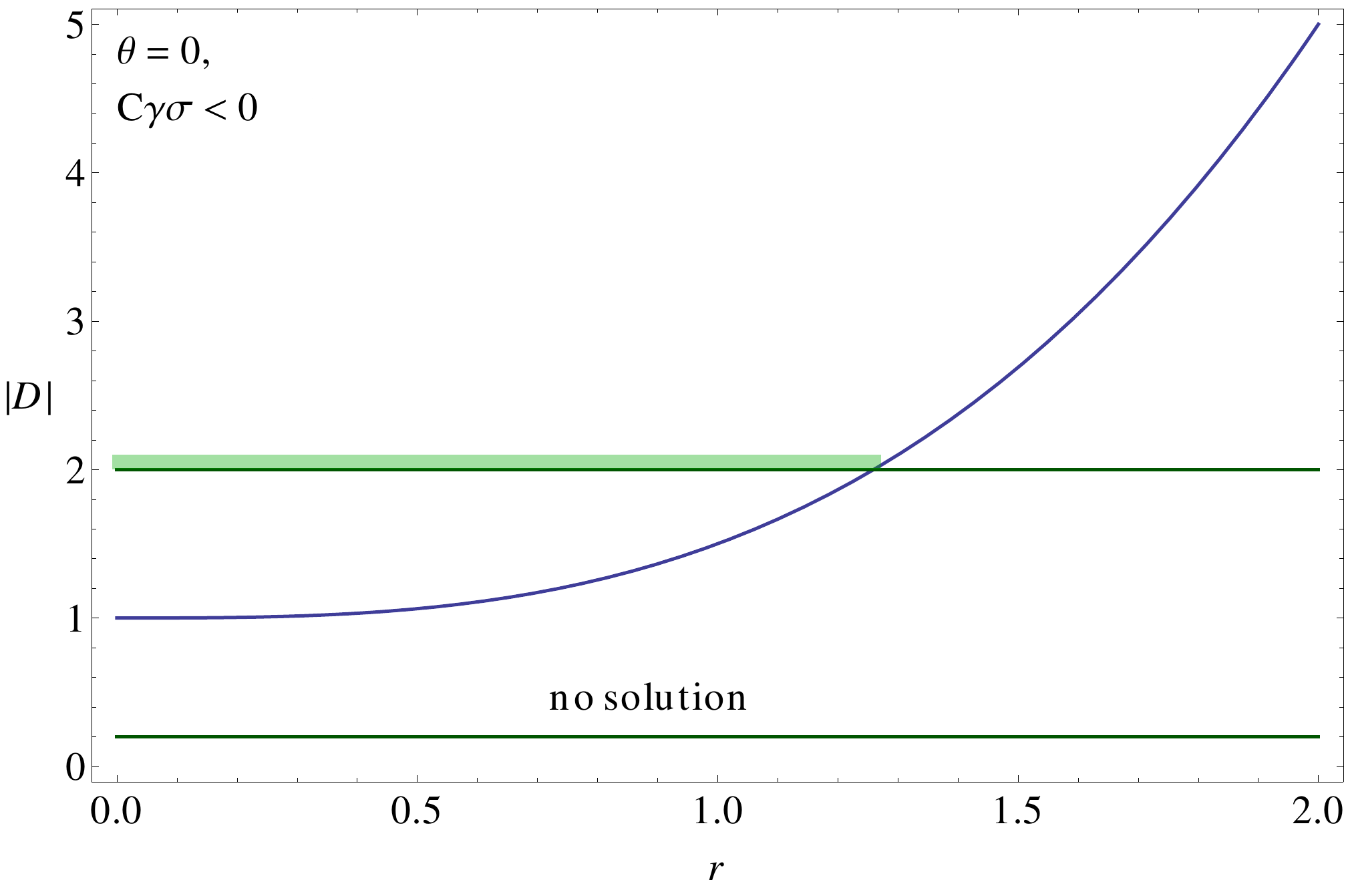}}
 \mbox{\hspace{0.01\textwidth}\includegraphics[width = 0.47\textwidth, height = 0.35\textwidth]{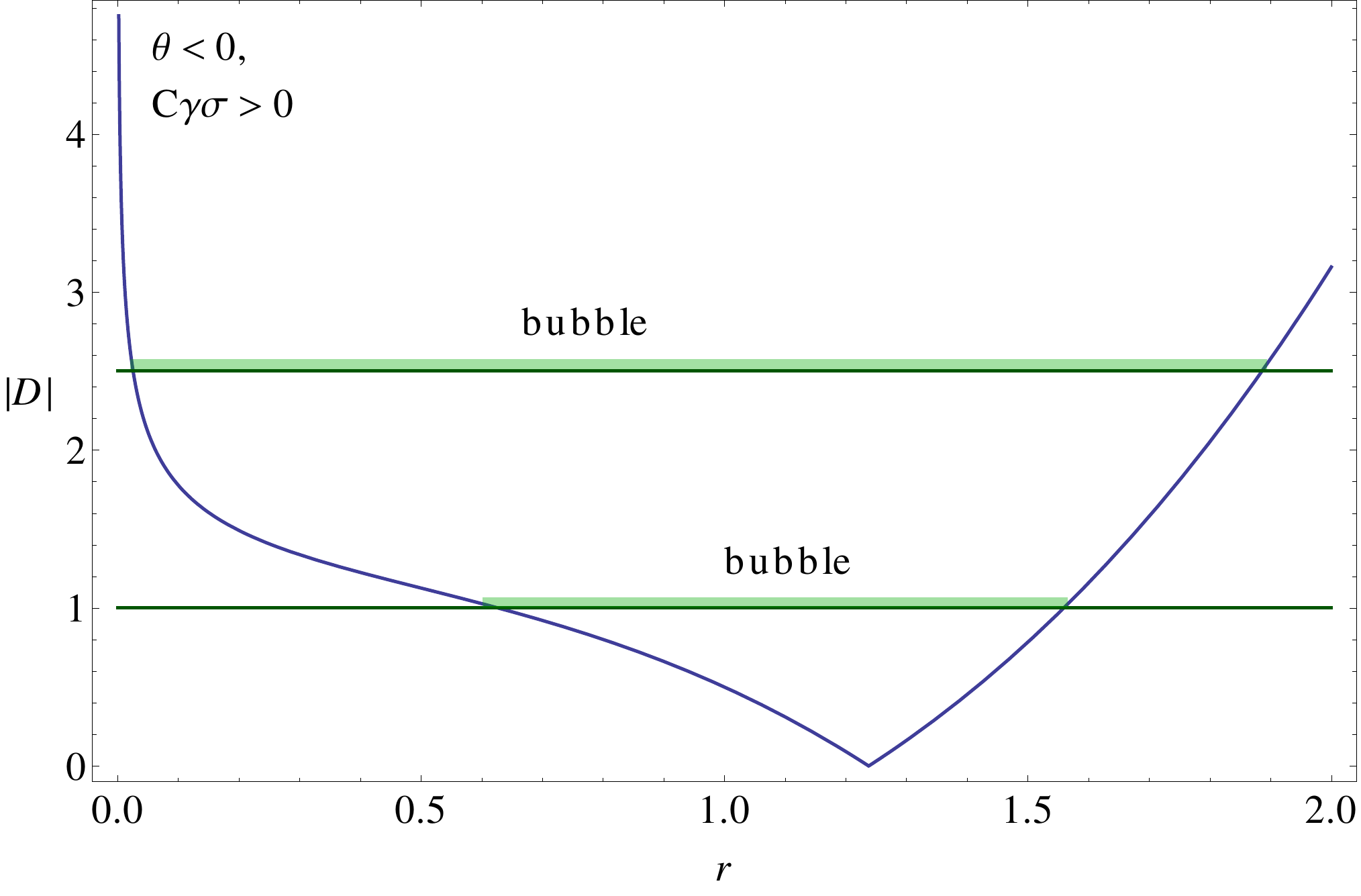}\hspace{0.005\textwidth} \includegraphics[width = 0.495\textwidth, height = 0.35\textwidth]{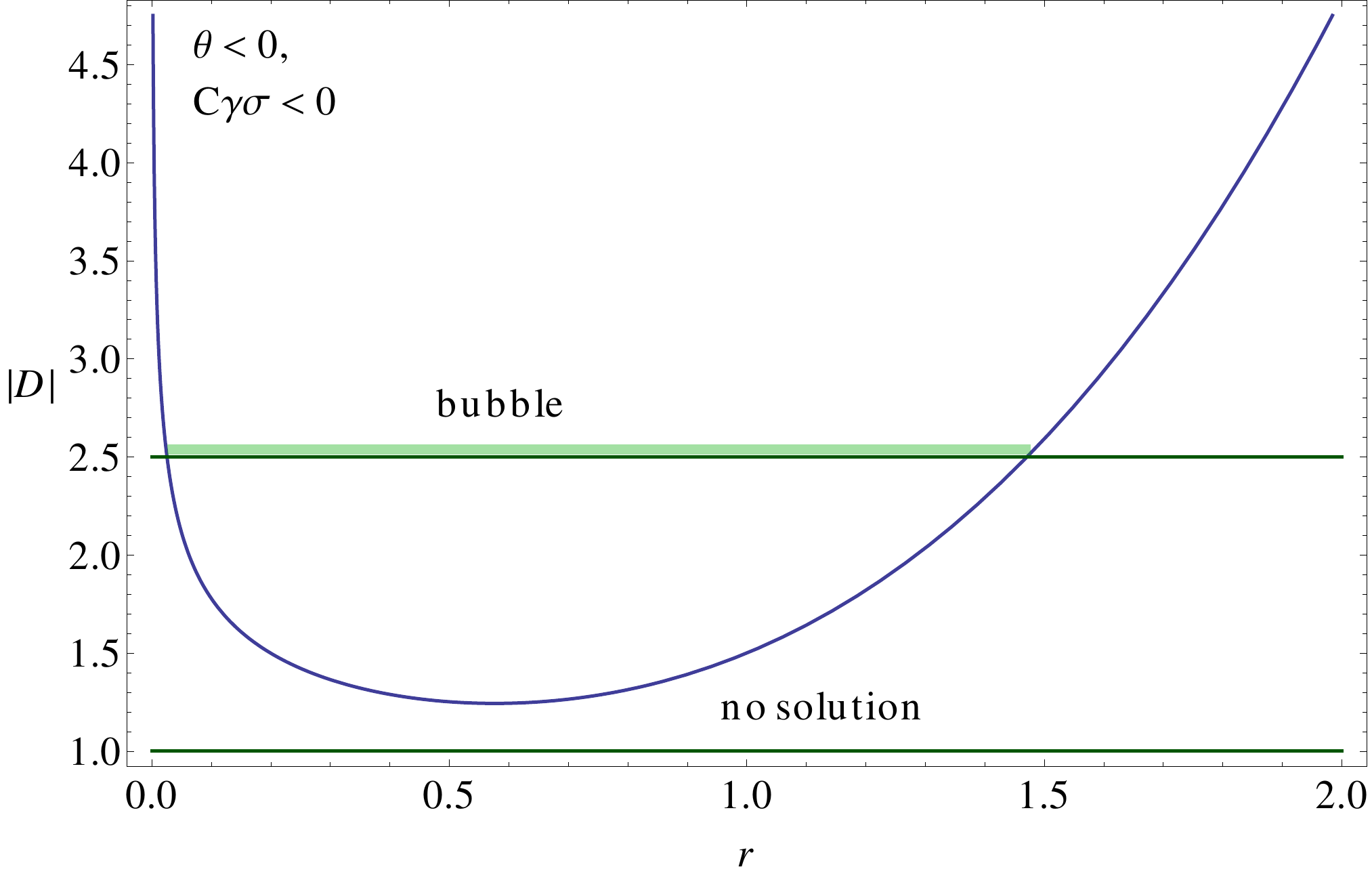}}
 \caption{Typical plots of the function $|D(r)|$ in various parts of parameter space. The horizontal lines correspond to different possible structures of solutions to $|D(r)| = 1/4G_N$. The shaded regions are where a solution to the Euler-Lagrange equation \eqref{eq:xdotsq} exists. Bubbles are solutions that exist without reaching the boundary.}
 \label{fig:plotD}
\end{figure}

We noted above that the surface has an endpoint at any radius $r_o$ at which $\dot x(r)$ diverges. From (\ref{eq:endd}), this occurs whenever
\be\label{eq:endpoint}
  |D(r_o)| = \frac1{4G_N}\, ; \quad  D(r) \equiv C r^d - \g \sigma\, r^{\theta/(d - \theta)}\, .
\ee
From (\ref{eq:xdotsq}) we can easily convince ourselves that this formula captures all endpoints.
As we have already mentioned, the structure of the above equation is such that multiple endpoints are possible. The  hypersurfaces $\G$ can consist of multiple disconnected hypersurfaces or of one hypersurface that does not reach all the way to the boundary. Both of these cases correspond to the formation of ``bubbles'' in the interior of the bulk. We will now describe the possible scenarios, with the plots in figure \ref{fig:plotD} below summarizing the various cases.

The behavior of $|D(r)|$ near the boundary is primarily determined by the sign of the exponent $\theta$. The fact that $\theta/(d - \theta) < d$ for any $\theta < d - 1$, implies that we will always have the scaling relation $|D(r)| \sim r^{\theta/(d - \theta)}$ at $r \rightarrow 0$. On the other hand, the large $r$ behavior will always be $|D(r)| \sim r^d$. A solution $x(r)$ will exist at those $r$ for which $|D(r_o)|\leq 1/4G_N$, and hence we can already conclude that any solution that exists at all will have at least one endpoint at a sufficiently large position $r_o$; there are no solutions that extend to $r = \infty$. We can also conclude that, if $\theta < 0$, any existing hypersurfaces will have an additional endpoint near the boundary. Thus, no solutions reach the boundary in geometries with $\theta < 0$.

Consider first $\theta < 0$. The function $|D(r)|$ decreases as we move away from the boundary, and at some point it will reach a minimum and then start growing again. See figure \ref{fig:plotD} below.
Thus, eq.~\eqref{eq:endpoint} will either never be satisfied or will be satisfied by two points $r_o$. In the former case, there is no solution. In the latter case, the hypersurface exists between the two endpoints, both with $r_o > 0$, and so it will form a ``bubble'' in the bulk.

If $\theta = 0$, the function $|D(r)|$ tends to $|\g \s|$ as we approach the boundary. The minimizing hypersurface will reach the boundary if $|\g \s| < 1/4G_N$. If the hypersurface does not reach the boundary, one may again find a bubble in the bulk. This occurs if $C\g\s > 0$; in this case $|D(r)|$ decreases as we move away from the boundary, vanishes at $Cr^d = \g\s$, and then increases with $r$ until it assumes the expected IR scaling as $r^d$. Eq.~\eqref{eq:endpoint} will then have two solutions corresponding to the two endpoints of the bubble. Alternatively, if $C \g \s < 0$ (while still having $|\g \s| > 1/4G_N$), there will be no solutions. See figure \ref{fig:plotD} below.

Finally, if $0 < \theta < d - 1$, there will always exist a hypersurface that satisfies the equation of motion and reaches the boundary. If $C \g \s < 0$, the function $|D(r)|$ will be monotonic, and it will intersect the horizontal line $D = 1/4G_N$ exactly once, at the point $r_o$ corresponding to the single endpoint of the hypersurface. If $C \g \s > 0$, similar reasoning to the $\theta = 0$ case leads to the conclusion that there are three endpoints. The solution therefore consists of two disconnected parts: one that reaches the boundary, and one forming a bubble. Again, see figure \ref{fig:plotD}.

The existence of bubble solutions may have interesting consequences for the quantum dynamics of our `dipole surfaces'.
It is of interest to return to these solutions once the microscopics of these surfaces is better understood. In terms of our immediate interest in solutions that reach the boundary, we see that no solutions exist at $\theta < 0$ or at $\theta = 0$ when the flux term dominates over the gravitational one, i.e.~when $|\g\s| > 1/4G_N$. In all other cases one can find values for the constant $C$ for which there exist admissible solutions. In particular, $C$ can be found such that the endpoint $r_o$ becomes large. As we noted in the previous subsection, these solutions have the property that $L = 2\int_0^{r_o} \dot x dr$ becomes large as $r_o$ becomes large. Generally, there may be more than one value of $C$, and hence more than one solution, for a given large $L$. In this cases, the surface with lowest deformed entropy should be chosen.\footnote{
A detailed examination of the candidate hypersurfaces reveals that some of them may have profiles whose derivatives $\dot x(r)$ change sign before diverging. The width of such hypersurfaces can increase as we go further into the IR, and sometimes they can exhibit self-intersections. Our numerical studies on specific bulk geometries have never resulted in one such surface being the global minimizer of the deformed entropy.}

For the cases in which there are no solutions that reach the boundary, one must consider more carefully the crossover between the IR scaling metrics we have been considering and the full spacetime, e.g.~an asymptotically AdS space. In general, though, one will still find situations in which, at large separations $L$, the asymptotic solutions are repelled from the near horizon region, to the extent that connected solutions do not exist at large $L$. This is similar to the behavior of minimal surfaces in confining geometries. In those case one finds disconnected solutions at large separation. In the case of the deformed entropy in the presence of a background charge density, however, it is not clear what the disconnected solutions would be, as the surface cannot make it all the way into the IR in order to cap off smoothly.

\section{Towards field theoretical order parameters}

Let us discuss separately the field theory interpretation of flux through minimal surfaces and of the action obtained by minimizing the deformed entropy.

Knowing the electric flux through a minimal Ryu-Takayanagi spatial hypersurface $\Gamma$ is equivalent, by Gauss's law, to knowing the electric charge contained within the surface and knowing the flux in the region at infinity delimited by the boundary $\partial \Gamma = \Sigma$ of the surface. See our discussion around figure \ref{fig:fluxgauss} above. The flux at infinity is just the total charge of the dual field theory contained within the region bounded by $\Sigma$, which is certainly a known quantity. It remains to describe the bulk charge contained within the bulk surface $\Gamma$ from a boundary perspective.

The physical significance of the bulk spatial region bounded by the bulk minimal surface $\Gamma$ is not well established. Nonetheless, it seems reasonable to expect that classical field configurations inside this bulk region are captured by the large $N$ reduced density matrix associated to the corresponding boundary region. This is the density matrix whose entanglement entropy is given by the area of the bulk minimal surface. Classical field configurations should include all data that can be computed by direct evaluation on the saddle point `master field' of the large $N$ gauge theory.\footnote{This perspective outlined in this paragraph has been emphasized to us by Matt Headrick. Related discussions concerning the region of the bulk captured by the reduced density matrix of a boundary region can be found in \cite{Czech:2012bh, Hubeny:2012wa}.} The bulk charge inside the surface is one such classical observable. Therefore, we may expect that the bulk charge we are interested in can be extracted from the large $N$ reduced density matrix. This would show, at least in principle, that the flux $\Phi_\Gamma$ is a well-defined field theory observable. However, it remains to 
be seen if 
a simple expression can be obtained for $\Phi_\Gamma$ in terms of the reduced density matrix of the region inside $\Sigma$.

We turn now to the interpretation of the deformed entropy as an action that should be extremized. As we noted in the previous section, this action describes a worldvolume theory with a tension and a dipole moment under the external electric field. To obtain the surface operator dual to such bulk worldvolume theories, they must be engineered (if possible) using D-branes in string theory. While polarization is commonplace among D-branes \cite{Myers:1999ps}, some work in a specific string theory background will be necessary to see if these effects can lead to the desired effective worldvolume actions. The essential question is how to maintain the charges on the (overall neutral) worldvolume separated, so that there is a net moment. An external electric field will of course polarize the worldvolume, but we are instead looking for an inherent dipole moment that can then interact with an external electric field.

From an effective field theory point of view we can ask what kind of worldvolume dynamics is necessary to generate or prevent the generation of a $\Phi_\Gamma \sim \int_\Gamma \star F$ term in the action. This term breaks charge conjugation $C$. Furthermore, the dipole moment induces a preferred orientation on the brane and therefore also breaks parity $P$. It follows that we can expect a dipole coupling to be present in the effective worldvolume action if the microscopic dynamics breaks $C$ and $P$ but preserves $CP$. This suggests that the surface operators we require could be dual to D-branes with chiral worldvolume theories.

A third way to get a handle on the field theory physics dual to the fractionalization-cohesion transitions we have discussed is by generalizing the large $N$ gauge theory approach to deconfinement introduced by \cite{Aharony:2003sx}. A gauge theory on a spatial sphere undergoes a Hagedorn transition at high temperatures due to the large growth in the number of long single trace operators. This transition describes deconfinement according to the Polyakov loop order parameter. Aspects of this story were generalized to finite density systems in \cite{Hollowood:2008gp}. The onset of fractionalization in large $N$ gauge theories at finite density should be describable in a similar language. Instead of the partition function, one should compute the charge density. In fractionalized phases the expectation value of the charge density will be dominated by long charged operators, while in a cohesive phase, the charge density will be carried predominantly by short charged operators.

\section*{Acknowledgements}

It is a pleasure to acknowledge helpful discussions with Matt Headrick, John McGreevy, Shiraz Minwalla, Subir Sachdev, Steve Shenker and Brian Swingle. The work of S.A.H.~is partially supported by a Sloan research fellowship. The work of \DJ.R.~is supported by an NSF Graduate Research Fellowship.

\end{document}